\documentclass[11pt,twoside]{article}

\usepackage{asp2006}
\usepackage{epsf}
\usepackage{lscape}

\begin{document}
\title{Gamma-rays from massive protostars}   
\author{Gustavo E. Romero \& Anabella T. Araudo} 
\affil{Instituto Argentino de Radioastronom{\'\i}a (CCT La Plata, CONICET), 
C.C.5, 1894 Villa Elisa,  Buenos Aires, Argentina\\
 Facultad de Ciencias Astron\'omicas y Geof\'{\i}sicas,
Universidad Nacional de La Plata, Paseo del Bosque, 1900 La Plata,
Argentina }  
\author{Valent{\'\i} Bosch-Ramon}   
\affil{Max Planck Institut f\"ur Kernphysik, Saupfercheckweg
1, Heidelberg 69117, Germany} 
\author{Josep M. Paredes}   
\affil{Departament d'Astronomia i Meteorologia and Institut de Ci\`encies 
del Cosmos (ICC), Universitat de Barcelona (UB/IEEC),  Mart\'{\i} i 
Franqu\`es 1, 08028, Barcelona, Spain}

\begin{abstract}
Massive protostars have associated bipolar outflows with velocities of 
hundreds of km s$^{-1}$. Such outflows produce
strong shocks when interact with the ambient medium leading to regions 
of non-thermal radio emission.
Under certain conditions, the population of 
relativistic particles accelerated at the terminal shocks 
of the protostellar jets can produce significant $\gamma$-ray emission. 
We estimate the conditions necessary for high-energy emission in the 
non-thermal hot spots of jets associated with massive protostars 
embedded in dense molecular clouds.  
Our results show that particle-matter interactions can lead to the 
detection of 
molecular clouds hosting massive young stellar objects by 
the {\it Fermi}{} satellite at MeV-GeV energies and even by Cherenkov
telescope arrays in the GeV-TeV range.
Astronomy at $\gamma$-rays can be used to probe the physical conditions in star 
forming regions and particle acceleration processes in the complex 
environment of massive molecular clouds.
\end{abstract}

\section{Introduction}  

Massive stars appear to be formed in the dense cores of massive cold 
clouds (Garay \& Lizano 1999, and references therein).  The
accumulation of gas in the core might proceed through previous stages
of fragmentation and coalescence progressively generating a
massive protostar that then accretes from the environment
(e.g. Bonnell, Bate \& Zinnecker 1997).
An alternative picture is direct 
accretion onto a central object of very high mass
(e.g. Rodr\'\i guez et al. 2008; see Shu et al.  1987 for the basic
mechanism). In any case, the prestellar core is expected to have
angular momentum which would lead to the formation of an accretion
disk. The strong magnetic fields inside the cloud that thread the disk
should be pulled toward the protostar and twisted by the rotation
giving rise to a magnetic tower, with the consequent outflows, as
shown by numerical simulations (e.g. Banerjee \& Pudritz 2007). 

Evidence supporting the existence of molecular outflows is found 
through methanol
masers, which are likely associated with shocks formed by the
interaction with the external medium (e.g. Plambeck \& Menten 1990).
However, the most important piece of evidence for outflows is in the form of
thermal radio jets. These jets are observed to propagate through the
cloud material along distances of a fraction of a parsec in some cases
(e.g. Mart{\'\i}  et al. 1993). At the end point of the jets, hot
spots due to the terminal shocks are observed in several sources. In a
few cases, these hot spots are clearly non-thermal, indicating the
presence of relativistic electrons that produce synchrotron radiation
(e.g. Araudo et al.  2007, 2008).

A population of relativistic particles in the complex environment of
the massive molecular cloud where the protostar is being formed will
produce high-energy radiation through a variety of processes: inverse
Compton (IC) up-scattering of IR photons from the cloud, relativistic
Bremsstrahlung, and, if protons are accelerated at the shock as well,
inelastic proton-proton ($pp$) collisions. If such a radiation is
detectable, $\gamma$-ray astronomy can be used to shed light on the star
forming process, the protostar environment, and cosmic ray
acceleration inside molecular clouds. In the present contribution we
want to discuss the conditions under what massive protostars can 
produce $\gamma$ rays detectable by the {\it Fermi} 
satellite in the immediate future,
and also whether Cherenkov telescopes may eventually detect these
sources. 
The models presented here incorporate many
refinements from the simpler approach published by Araudo et
al. (2007).

\section{Basic scenario}
\label{sce}

The basic scenario has been outlined in Fig. 1 of Araudo et al. (2007).
A massive young stellar object (YSO), or a group of them,
is embedded into a molecular cloud. The protostar heats the
cloud in such a way that it can be detected as a strong IR source,
with luminosities in the range  $L_{\rm IR}\sim 10^{4-5}$ $L_{\odot}$
$\sim10^{38-39}$ erg s$^{-1}$ whereas the optical counterpart is obscured by
the cloud. Masses and sizes of these clouds are of the order of $\sim
10^{3}$~$M_{\odot}$ and few pc, respectively (e.g. Garay \& Lizano
1999).  The thermal jets can propagate up to distances of $10^{16-18}$
cm from the central source, and the non-thermal radio lobes are seen
at distances of $\sim$ pc from the protostar (Mart{\'\i}  et al. 1993, 
Garay et al. 2003) with sizes of the order of 1
\% this distance. The radio lobes are ionised by the terminal shocks
and particles can be accelerated up to relativistic energies both in
the reverse shock, as well as in the forward shock in the ionised
gas. In what follows we will discuss the characteristics of the
non-thermal particle population of the lobes arose from forward and reverse 
shocks.

A crucial parameter for the acceleration of particles up to high
energies is the magnetic field $B$. Equipartition calculations with the
non-thermal population of particles (Araudo et al. 2007) and Zeeman 
measurements
(e.g. Crutcher 1999) suggest $B \sim 1$~mG. Magnetic
fields are thought to play an important role supporting the cloud
before the gravitational collapse, allowing high densities in the
cores to be achieved (e.g. McKee \& Ostriker 2007). The medium density
strongly determines the jet evolution, and is a key factor regarding
the emission at high energies.

Here, we consider the effect of both a
forward and a reverse shock, the escape of the particles from the
acceleration region, and the illumination of the whole molecular cloud
by protons that diffuse from the lobes,
among others refinements with respect to our previous work. 
The cloud acts as
reservoir of locally accelerated cosmic rays. In very massive clouds,
where several protostars are forming, this effect can be very
significative.

\section{Particle acceleration and losses}

Diffusive shock acceleration (DSA, e.g. Drury 1983)
can operate in the turbulent region at the termination of
the thermal jets. If the jet head is still moving in the free
propagation phase, the bow shock in the cloud can
be as fast as the jet, but then the shock luminosity must be smaller
than the one carried by the jet ($L_{\rm j}$). In any case, regardless
the energy budget, the fast velocities allow particle acceleration up
to very high energies. 
On the other hand, when the swept cloud material is massive enough, the jet
slows down and the bow shock produced by the interaction can be as
slow as $\sim 10$~km~s$^{-1}$. The shocked material of the cloud will be 
radiative,
forming a very thin shell pushed by the jet. This shock is not
suitable to accelerate particles, but its downstream material 
would be a good target for
$pp$ interactions or relativistic Bremsstrahlung in case that
electrons and protons accelerated elsewhere could reach it. In the
slow bow-shock case, the region suitable for particle acceleration is the
reverse shock formed in the jet, which was not strong when the bow
shock was moving as fast as the jet itself (being an effective
accelerator). When the bow shock has slowed significantly down, the
reverse shock has a velocity $v_{\rm s}\sim v_{\rm jet}$. The shock
luminosity in the slow bow-shock case is similar to the jet one.

In any case (bow shock or reverse shock acceleration), the particle 
acceleration rate for a parallel
non-relativistic strong shock in the test particle approximation is
 
\begin{equation}
\left.\frac{dE_{e,p}}{dt}\right|_{\rm gain}\sim \eta qB c, \qquad 
\eta=\frac{3}{20f_{\rm sc}}\left(\frac{v_{\rm s}}{c}\right)^2,
\end{equation}
where $e$ and $p$ stand for electrons and protons,
respectively (e.g. Protheroe 1999). The parameter
$f_{\rm sc}$ is the ratio of the mean free path of the particles
to their gyro-radius $r_{\rm g}$. Close to the Bohm limit, where the
diffusion coefficient is $D_{\rm B}=c r_{\rm g}/3$, $f_{\rm sc}\sim 1$. For
outflows with bulk velocities $v_{\rm jet}\sim 10^3$~km~s$^{-1}$ (e.g.
Mart{\'\i} et al. 1995) we get a relatively low efficiency of
$\eta\sim 10^{-6}$, which is, nonetheless, enough to achieve
relativistic energies for the particles. 

\subsection{Hadronic losses}

The relevant energy loss for protons is due to $pp$ collisions with
the nuclei of the ambient gas of density $n$:

\begin{equation}
\left.\frac{dE_{p}}{dt}\right|_{pp} = n\, c\, \sigma_{pp}(E_p) K_{pp} E_{p}, 
\end{equation}
where
$\sigma_{pp}(E_p) = 34.3 + 1.88L + 0.25L^2 \;\;\rm{mb}$, with $L =
\ln(E_p/1\;\rm{TeV})$, is the interaction cross section (Kelner at
al. 2006) and $K_{pp}\sim0.5$  the inelasticity. The emission from
$pp$ interactions can be  significant from high- to very high-energy
$\gamma$ rays.

In addition to the radiative losses, protons can escape from the
emitting region of size $l_{\rm lobe}$ by diffusion or convection
downstream the shock,  in accordance with the timescales of these
processes: $t_{\rm diff}\sim l_{\rm lobe}^{2}/2\,D_{\rm B}$ and $t_{\rm
  conv}\sim 4\,l_{\rm lobe}/v_{\rm s}$. The escape time $t_{\rm esc}$ 
is defined as the minimum between $t_{\rm diff}$ and $t_{\rm conv}$.
In the presence of a turbulent $B\sim 1$ mG or higher, $t_{\rm conv}$ will
usually be shorter than $t_{\rm diff}$ and 
$t_{pp} = E_{p} (dE_{p}/dt)_{pp}^{-1}$. 
Nevertheless, in
the cases studied here the most energetic protons have their energy
constrained by diffusive escape upstream. 

\subsection{Leptonic losses}

The main losses for electrons are due to synchrotron radiation, IC
up-scattering of IR photons, and relativistic Bremsstrahlung. The
standard formulae are presented, for instance, in Blumenthal \& Gould
(1970). Synchrotron and IC losses in the Thomson regime satisfy:
\begin{equation}
\left(\left.\frac{dE_e}{dt}\right|_{\rm synch}\right)
\left(\left.\frac{dE_e}{dt}\right|_{\rm IC}\right)^{-1}
=\frac{u_B}{u_{\rm ph}} > 1 ,
\end{equation}
where $u_{B}=B^{2}/8\pi$ is the magnetic energy density ($\sim
10^{-8}$~erg~cm$^{-3}$ for $B=1$~mG), and $u_{\rm ph}$  is the photon
energy density inside the cloud, typically $u_{\rm ph}\sim
10^{-9}$~erg~cm$^{-3}$, which is dominated by IR photons.  This means
that synchrotron losses will dominate over IC cooling. The highest
energy electrons will cool via synchrotron and they will radiate from
radio to soft X-rays through this mechanism.  On the other hand,
lower energy electrons are dominated by relativistic  Bremsstrahlung
and escape losses from the emitter.  The relativistic Bremsstrahlung
component contributes to the spectrum from hard X-rays up to high
energies.

For very high densities, ionization/Coulomb collision losses may
dominate radio emitting electrons, hardening their spectra.  An
observed steep radio spectrum implies that this cooling channel is
minor in comparison to synchrotron radiation.

\subsection{Maximum energies and particle spectra}

In the Bohm limit we expect for electrons that the maximum energy will
be determined by the radiative losses. Equating energy gains and
losses we get, for the parameters of the basic scenario sketched in
Sect. \ref{sce}, that $E^{\rm max}_{e}\sim 1$ TeV. For protons, on the
contrary, cooling times are longer and hence they can diffuse away
from the shock upstream into the cloud. The residence time in the
shock region, of size $\sim l_{\rm lobe}$,
determines the maximum energy: $E^{\rm max}_{p}\sim 1-100$ TeV.

The particle distributions $N_{e,p}(t,E_{e,p})$ are given by the transport
equation (e.g. Ginzburg \& Syrovatskii 1964):
\begin{equation}
\frac{\partial N_{e,p}\left(t,E_{e,p}\right)}{\partial t}+\frac{\partial
  \left[b_{e,p}(E_{e,p})
    N_{e,p}\left(t,E_{e,p}\right)\right]}{\partial E_{e,p}}+\frac{N_{e,p}
\left(t,E_{e,p}\right)}{t_{\rm esc}}=Q_{e,p}(t,E_{e,p}),
\label{Ginz_equation}
\end{equation}
where $Q_{e,p}(t,E_{e,p}) \propto E_{e,p}^{-\Gamma_{e,p}}$ is the function for the 
particle injection (a power law according to the DSA mechanism which is 
supposed to operate
in the radio lobes). Injection takes place during the lifetime of the
source ($t_{\rm life}$), estimated typically in $t_{\rm life}\sim
100-1000$~yr $\approx 3\times (10^9-10^{10})$~s  for massive O
protostars (see Garay et~al. 2003). In the transport equation $t$ is
time, and $b_{e,p}(E_{e,p})$ is the summation of all energy losses. 

Eq. (\ref{Ginz_equation}) can be solved in the steady state
setting $\partial N_{e,p}\left(t,E_{e,p}\right)/\partial t=0$.  In the
cases when synchrotron dominates over escape or relativistic
Bremsstrahlung, the electron distribution will be
$N_e(E_{e})\propto Q_e E_e^{-1}$ at high energies.  Relativistic 
Bremsstrahlung losses and escape, on
the contrary, lead to $N_e(E_e)\propto Q_e(E_e)$. The injected
proton spectrum is not modified since the relevant losses, due to
inelastic $pp$ collisions, are almost linear with the energy in the
range considered (from $\sim$ 100~MeV to TeV) $N_p(E_{p})\propto Q_{p}$.

The normalization of the electron spectrum is given by the observed
radio luminosities, between $10^{28}$ and $10^{30}$~erg~s$^{-1}$ for
sources like HH 80-81 (Mart{\'\i} et al. 1993) and IRAS~16547$-$4247
(Rodr{\'\i}guez et al.  2005). The proton to electron energy density 
ratio, $a = u_{p}/u_{e}$, cannot be established from first principles in
the acceleration process 
and it should be considered as a phenomenological parameter. All this
fixes the luminosity injected in non-thermal particles  ($L_{\rm
  nt}$).  In the next section we will consider two cases: one with
only electrons ($a=0$) and another dominated by protons, with a ratio
$a=100$, as observed in galactic cosmic rays. 

\section{Non-thermal spectral energy distributions}

Once the particle energy distribution $N_{e,p}(E_{e,p})$ has been obtained 
through 
Eq. (\ref{Ginz_equation}), the spectral energy distribution
(SED) can be calculated as:
\begin{equation}
E_{\gamma}L_{E_{\gamma}}=E_{\gamma}^{2}\int q_{\gamma}(E_{\gamma},E_{e,p}) 
 N_{e,p}(E_{e,p}) dE_{e,p},
\end{equation}
where $q_{\gamma}$ is the particle emissivity of the particular 
radiation process and $L_{E_{\gamma}}$ is the specific luminosity. 

\subsection{IRAS~16547$-$4247}

The powerful IR source IRAS~16547$-$4247 is composed by a massive 
protostar embedded
in a very dense and luminous molecular cloud (Garay
et al. 2003). In addition, two radio lobes are at the end of the 
thermal jets of the central YSO, being the south one clearly non-thermal. 

In Fig.  \ref{SED} (\emph{Left}) we show an example of the
SED for the non-thermal lobe of the IRAS source, where only primary electrons
have been considered.  In this specific case, the reverse shock is
accelerating particles  since the cloud material is very dense 
and the bow shock has been already slowed down.
Densities have been taken as high as possible accounting for the
ionization cooling limit.  The parameters used in the
modeling are presented in Table~\ref{tab}.  This example shows that,
despite the IC contribution is negligible, $\gamma$-rays might still be
of leptonic origin if densities are particularly  high. The spectral
shape  is determined by the escape of electrons from the emitter (this
was not considered in Araudo et al. 2007).  The SED has two clear
peaks, one due to synchrotron radiation and the second one due to
relativistic Bremsstrahlung. The latter emission extends into the
$\gamma$-ray domain with integrated luminosities of  $\sim 5\times
10^{32}$ and $3\times 10^{31}$~erg~s$^{-1}$ at $E_{\gamma}\sim
0.1-100$ and $>100$~GeV, respectively.  
The emission for the case with $a=100$ has been also estimated but not shown
in the present paper. 
Synchrotron emission is now dominated by secondary pairs,  and the
$pp$ in the lobe reaches integrated  luminosities of $\sim 2\times
10^{33}$ and  $2\times 10^{32}$~erg~s$^{-1}$ at $E_{\gamma}\sim
0.1-100$ and $>100$~GeV, respectively. 

\subsection{HH~80-81 system}

The Herbig-Haro objects HH~80 and HH~81 are located at the terminal point 
of a massive YSO jet. The triple system  composed by the central YSO plus
HH~80-81 and HH~80 North is at the edge of a dark cloud 
(Mart\'i et al. 1993).

\begin{figure}[!h]
\plottwo{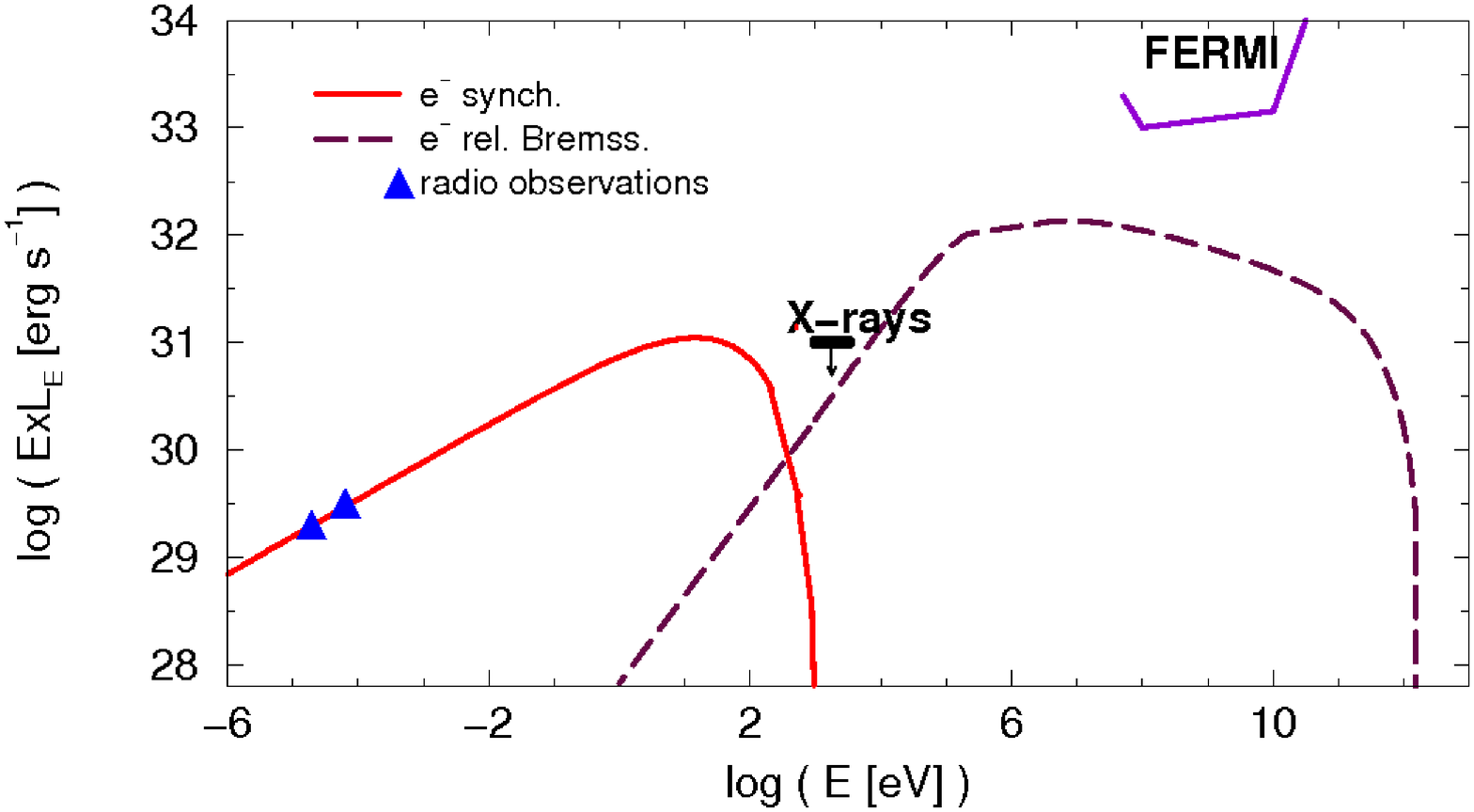}{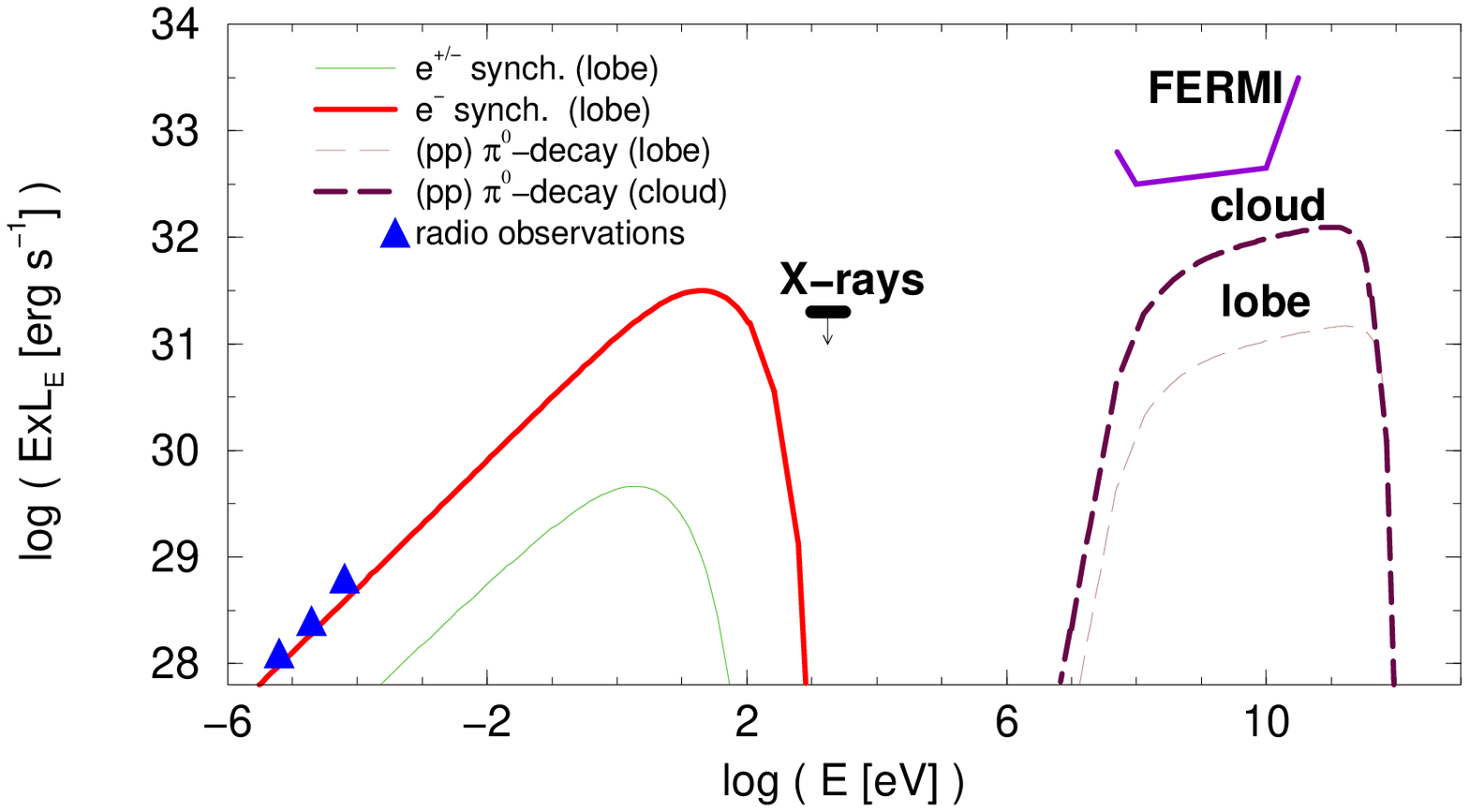}
\caption{
{\itshape Left:\/} 
SED of the south lobe of the source IRAS~16547$-$4247 for a pure leptonic 
case (a=0).
The IC contribution is negligible and not shown here. 
Radio and X-ray observational points are from Rodr{\'\i}guez 
et al. (2005) and from Araudo et al. (2007), respectively. 
{\itshape Right:\/}
SED of the source HH~80 for a case dominated by protons 
($a=100$). The synchrotron radiation is dominated by primary electrons.
Radio and X-ray observational points are from Mart{\'\i} et al. (1993) and 
Pravdo, Tsuboi \& Maeda (2004), respectively.
Both the emission from the lobe and that which results from 
the proton diffusion in the cloud are shown.
In both figures  we indicate the
sensitivity curve of {\it Fermi} for 1~yr of observations and the
distance of the sources, $\sim 3$ and 1.7 kpc for  IRAS~16547$-$4247
and HH~80-81, respectively.
}\label{SED}
\end{figure}

In Fig. \ref{SED} (\emph{Right}), we show a
case dominated by protons, with $a=100$, for the source HH~80.
Given the lower densities of the environment,
the bow shock could still  move fast and therefore be an efficient
particle accelerator.  In this source the radio emission is still
produced by primary electrons, with some contribution from secondary
pairs injected through the decay of charged pions. At high energies
the IC and the relativistic Bremsstrahlung are negligible and the
emission is dominated by $\gamma$ rays  originated in neutral pion
decays from $pp$ interactions. The $\gamma$-ray cutoff appears at 
energies $E^{\rm
  max}_{\gamma}\sim  E^{\rm max}_{p}/10$. The total $\gamma$-ray
integrated luminosities from the lobe can reach $\sim
10^{32}$~erg~s$^{-1}$.  However, since protons can diffuse out the
acceleration region with significant energy due to their long cooling
times, the interaction with the cold material of the cloud will add a
diffuse component (Aharonian \& Atoyan 1996). This emission could look
marginally  extended for current Cherenkov telescopes and {\it
  Fermi}. The diffuse radiation from the massive molecular cloud can
increase the $\gamma$-ray integrated luminosity to the level of $\sim
5\times 10^{32}$ and  $6\times 10^{32}$~erg~s$^{-1}$ at
$E_{\gamma}\sim 0.1-100$ and $>100$~GeV, respectively, rendering the
source detectable. 
The leptonic case (a=0) for the object HH~80 was also computed but the
luminosities at high energies are  well below $10^{30}$~erg~s$^{-1}$
and the spectra are not shown here.

\begin{table}[!ht]
\caption{Physical parameters considered in this work.}
\label{tab}
\begin{center}
\begin{tabular}{l|c|cc}
\hline\noalign{\smallskip}
{} &  IRAS 16547-4247 & HH~80 & Cloud\\
\hline\noalign{\smallskip}
$l_{\rm lobe}$ [cm] & $1.1\times10^{16}$ & $5\times10^{16}$&{}\\
$L_{\rm j}$ [erg~s$^{-1}$] & $10^{36}$ & $1.5\times 10^{36}$&{} \\
$t_{\rm dyn}\,^a$ [s] &$4\times 10^8$ & $3\times 10^9$&$10^{11}$ \\
\hline
\hline
$a$ & 0 & 100&{}\\
$n$ [cm$^{-3}$] & $10^6$& $10^3$& $2.5\times 10^2$ \\
$B$ [mG] & 1 & 0.3 &{}\\
\hline
particle & $e$  & $p$ & {}\\
\hline
$L_{\rm nt}/L_{\rm j}$ & 0.02 & 0.1 &{}\\
$\Gamma$ & 2.2 & 1.8 &{}\\
\noalign{\smallskip}
\hline
\end{tabular}
\end{center}
{
$^{a}$ Dominant timescale: it is approximately $t_{\rm life}$ for 
HH~80 and the cloud, and $t_{\rm esc}$ for the IRAS source.} 
\end{table}

\section{Summary and discussion}\label{disc}

In the present contribution, we study the production of
$\gamma$-ray emission at the terminal points of jets that emanate
from massive protostars. 
In particular, we consider two sources:
IRAS~16547-4247 and HH~80-81. In the former case, ralativistic particles 
are accelerated in the shocked jet, wereas in the latter, in the bow shock. 
With the model presented here, $\gamma$-ray luminosities up to 
$\sim 10^{32}$~erg~s$^{-1}$ are obtained.

As pointed out by Romero (2008), the detection of massive protostars
at $\gamma$-ray energies would open a new window to star formation
studies. The detection of the cutoff in the SED will give important
insights on the acceleration efficiency in the terminal shocks of the
outflows. This, in turn, can be used to constrain the shock velocity
and the diffusion coefficient.  The accumulation of cosmic rays
accelerated in the radio lobes into the molecular cloud can produce
extended $\gamma$-ray sources. 

The combined effect of several protostars
deeply embedded in giant clouds might be responsible for GeV-TeV
sources found in star forming regions by EGRET, {\it Fermi}, {\it
  AGILE} and Cherenkov telescopes.  
A clustering of gamma-ray sources
should be present in regions with large molecular clouds and star
formation, as already inferred from EGRET data  (e.g.  Romero et
al. 1999).  Although we do not expect that massive protostars should
be among the bright sources detected by {\it Fermi} (Abdo et
al. 2009), our predictions show that they should show up in further
analysis of weaker sources after few years of observations. The
emission levels above 100~GeV, of about 0.01~Crab, could be detectable
by current and future Cherenkov telescopes for $\sim 50$~hours
observation time. 

Cosmic-ray re-acceleration inside the clouds due to
magnetic turbulence (e.g. Dogiel et al. 2004) could result in stronger
sources. Neither UV nor hard X-ray counterparts related to thermal
Bremsstrahlung produced in the shock downstream regions are expected
to be observed from these sources because of the large absorption
and/or low emission levels. In any case, massive clouds with high IR
luminosities and maser emission (tracers of massive star formation)
deserve detailed study with {\it Fermi} and other $\gamma$-ray
telescopes.

\acknowledgements 
G.E.R. and A.T.A. are supported by the Argentine agency ANPCyT.
G.E.R., V.B-R., and J.M.P  acknowledge support by the Ministerio de 
Educaci\'on y Ciencia (Spain) 
under grant AYA 2007-68034-C03-01, FEDER funds. A.T.A. thanks Max 
Planck Institut fuer Kernphysik for his kind hospitality 
and suport. 
V.B-R. gratefully acknowledges support from the Alexander von Humboldt
Foundation. The authors thank very much Josep Mart{\'\i} for organizing an 
excelent and enjoyable workshop.


\begin{thebibliography}{}
\bibitem{} Abdo, A.A., et al., 2009, arXiv0902.1340
\bibitem{} Aharonian, F.A., Atoyan, A.M., 1996, A\&A, 309, 917 
\bibitem{} Araudo, A.T., Romero, G.E., Bosch-Ramon, V., Paredes, J.~M., 
2007, A\&A, 476, 1289
\bibitem{} Araudo, A.T., Romero, G.E., Bosch-Ramon, V., Paredes, J.~M., 
2008, IJMP D, 17, 1889
\bibitem{}Banerjee, R., Pudritz, R.E., 2007, ApJ, 660, 479
\bibitem{} Blumenthal, G.R., Gould, R.J., 1970, Rev. Mod. Phys., 42, 237
\bibitem{} Bonnell, I.A., Bate, M.R., Zinnecker, H., 1998, MNRAS, 298, 93
\bibitem{} Crutcher, R.M., 1999, ApJ, 520, 706
\bibitem{} Dogiel, V.A., Gurevich, A.V., Istomin, Ya. N., Zybin, K.P., 
2005, Ap\&SS, 297, 201
\bibitem{} Drury, L.O'C., 1983, Reports on Progress in Physics, 46, 973
\bibitem{} Garay, G. and Lizano, S., 1999, PASP, 111, 1049
\bibitem{} Garay, G., Brooks, K., Mardones, D., Norris, R.P., 
2003, ApJ, 537, 739
\bibitem{} Ginzburg, V.L., Syrovatskii, S.I., 1964, The Origin of 
Cosmic Rays, Pergamon Press, New York
\bibitem{} Kelner, S.R., Aharonian, F.A., \& Vugayov, V.V., 2006, 
Phys. Rev. D, 74, 034018
\bibitem{} Mart{\'\i}, J., Rodr{\'\i}guez, L.F., Reipurth, B., 1993, 
ApJ, 416, 208
\bibitem{} Mart{\'\i}, J., Rodr{\'\i}guez, L.F., Reipurth, B., 1995, 
ApJ, 449, 184
\bibitem{} McKee, C.F., Ostriker, E.C., 2007, ARA\&A, 45, 565
\bibitem{} Plambeck, R.L., Menten, K.M., 1990, ApJ, 364, 555
\bibitem{}  Pravdo, S.~H., Tsuboi, Y., Maeda, Y. 2004, ApJ, 605, 259
\bibitem{} Protheroe, R.J., 1999, in: Topics in Cosmic-Ray Astrophysics, 
1999, p.247 [astro-ph/9812055]
\bibitem{} Rodr{\'\i}guez, L.F., Garay, G., Brooks, K., Mardones, D., 
2005, ApJ 626, 953
\bibitem{} Rodr{\'\i}guez, L.F., Moran, J.M., Franco-Hern\'andez, R, 
et al., 2008, AJ 135, 2370
\bibitem{} Romero, G.~E., Benaglia, P. \& Torres, D.~F. 1999, A\&A, 348, 868
\bibitem{} Romero, G.~E., 2008, in: High-Energy Gamma-Ray Astronomy, 
eds. Aharonian et al., AIP Conf. Proc., Vol. 1085, p. 97 
\bibitem{} Shu, F.H, Adams, F.C., Lizano, S. 1987, ARA\&A, 25, 23 
\end{thebibliography}
\end{document}